\documentstyle[12pt]{elsart}
\newcommand{\sixteenms}{16.$\bar{6}$~ms~}
\begin{document}

\begin{frontmatter}
\title{An electronic clock for correlated noise corrections.}
\author{W.J. Llope,$^\dag$ N. Adams, and K.K. Kainz}
\address{T.W. Bonner Nuclear Laboratory, \\
Rice University, Houston, TX 77005-1892}
\begin{abstract}
An inexpensive and portable approach to measure the time an
experimental event occurs as measured by a specific electronic clock
is presented. The clock resets in active synchronization with the
experimental AC-power cycle. This allows an efficient and complete
correction for correlated noise contributions to pulse area and time
measurements of detector channels equipped with PhotoMultiplier Tubes.
The electronic board that was developed will be described. The
performance for the treatment of correlated noise in experimental data
taken at the BNL-AGS facility, and analyses of spectral decompositions
of this noise, will also be described.
\end{abstract}
\begin{keyword}
noise, correlated, common mode, AC line, ground loop, ADC
pedestal, 60 Hz, synchronized electronic clock
\end{keyword}
\end{frontmatter}
PACS Numbers: 84.30.Ng, 7.50.-e, 89.20.+a

\section{Introduction\label{sec:intro}}

Despite dedicated efforts to insure otherwise, a voltage ripple with
the same frequency as the main AC power line (60 Hz in the U.S.) but a
complicated shape and amplitude may exist on the electrical ground
used in an experiment. The cause of the ripple is the existence
somewhere in the experiment of at least one electrical path between
the experimental ground and other grounds. These ``dirty" grounds may
be at different mean voltages relative to the experimental ground and may
be correlated with the AC power with different phases, line
shapes and amplitudes. Generally, experimenters use oscilloscopes and
voltmeters to locate unwanted electrical connection(s) in the
experiment that tie the experimental ground to dirty grounds. The most
egregious paths can generally be located and removed in this manner,
and the grounds may hence appear very clean when only the experiment
itself is fully powered up. During an actual run, however, the
experimental ground may be compromised in ways beyond the control of
the experimenters. The contaminating voltage ripple is a noise current
which may differ for different detectors in the experiment and for
different elements of any given detector, and may be weakly time
dependent over periods of hours. The digitization in an Analog-Digital
Converter (ADC) of the total charge of electrical pulses from active
detector elements is then made versus a baseline ground that itself
carries current. This smears ADC data, possibly significantly, and
makes less significant all quantities inferred from it. The custom
circuit developed to address this problem for Experiment 896
\cite{ref:e896} at the BNL-AGS, and its performance for correlated
noise corrections, is described here.

Detector channels involving PhotoMultiplier Tubes (PMTs) are
particularly sensitive to such ground contaminations, as PMTs
are nearly perfect amplifiers. Depending on certain
factors such as the widths of ADC gates that are used, even a few
millivolts of correlated noise can significantly worsen the ADC
resolution both on the pedestal and for hits. 
The Time of Flight (TOF) system in BNL-AGS Experiment 896
is composed of $\sim$190 Bicron BC404 plastic scintillator slats of various dimensions
each read out by two PMTs (Hamamatsu R2076 or 1398 depending on the slat). 
The signals from the E896 TOF System have a width of $\sim$10 ns FWHM
and peak at about -300 mV for single hits of relativistic charge
$Z$$=$1 particles. This implies that the integrated charge in a typical PMT
pulse for a $Z$$=$1 hit is on the order of 45 pC into 50 $\Omega$. For
this system, the ADC gate width used is 200 ns wide. If the ripple in
the ground at the time of a particle hit is 2 mV, the integral
within the same 200 ns wide gate is $\sim$8 pC into 50 $\Omega$, which
is $\sim$10\% of the charge measured for hits and {\it time
dependent}. Depending on how ADC information from a particular
detector is used in subsequent analyses, correlated noise may thus
have far reaching implications. The resolution on the charge $|Z|$/e
of particles striking (thin) E896 TOF slats, inferred from the TOF ADC
values, can be worsened. Corrections to the TOF timing information
based on the ADC information, {\it i.e.} ``slewing"
corrections,\cite{ref:slew} would be degraded, worsening the TOF
timing performance. There are other detectors in E896 that include the
PMT read-out of scintillation or \^{C}erenkov light. These detectors
include the Multi-Functional Neutron Spectrometer (MUFFINS), the Beam
Counters (BCs), the Exit Charge Detector (ECD), and the MuLTiplicity
detector (MLT). Correlated noise degrades $|Z|$/e measurements and the
quality of the slewing corrections for the MUFFINs and BC data, and
degrades inferences on the event centrality based on the MLT and ECD
data.

This paper is organized as follows. Section \ref{sec:symptom}
describes the apparent symptoms of a correlated noise problem, using
as an example those in experimental data collected by the E896
Collaboration \cite{ref:e896} during the Spring 1998 run of 11.5
GeV/c/N $^{197}$Au beams at the BNL-AGS. As alluded to above, the E896
experiment is composed of many different subsystems arranged in a
mechanically and electrically complicated configuration. There are two
large spectroscopic magnets in E896, and many more beam line magnets
nearby. As will be shown, correlated noise is rampant in E896 during
full beam-on running conditions, although the experimental grounds are
clean otherwise. The general solution chosen to combat this problem
offline involves custom electronics which are described Section
\ref{sec:blackbox}. Sections \ref{sec:results} and \ref{sec:fourier}
describe the results obtained when this information was inserted into
the E896 data stream during full beam-on running conditions. The
summary and conclusions are presented in Section \ref{sec:summary}.

\section{The Symptom\label{sec:symptom}}

In general, an ADC pedestal for a PMT-equipped detector channel
has contributions from the intrinsic offsets of the ADC itself,
and the PMT's dark current. For the Hamamatsu R2076 and 1398 PMTs
and the LeCroy 1885F ADCs used in the E896 TOF System,
these contributions typically result in a
pedestal variance on the order of 3-5 ADC channels.
A correlation between the pedestals of different detector channels
is the unambiguous signature of correlated noise, also known
as ``common mode noise," or ``ground loops." 

\begin{figure}[htb]
\vspace{8cm} 
\hspace{3cm}
\special{epsf=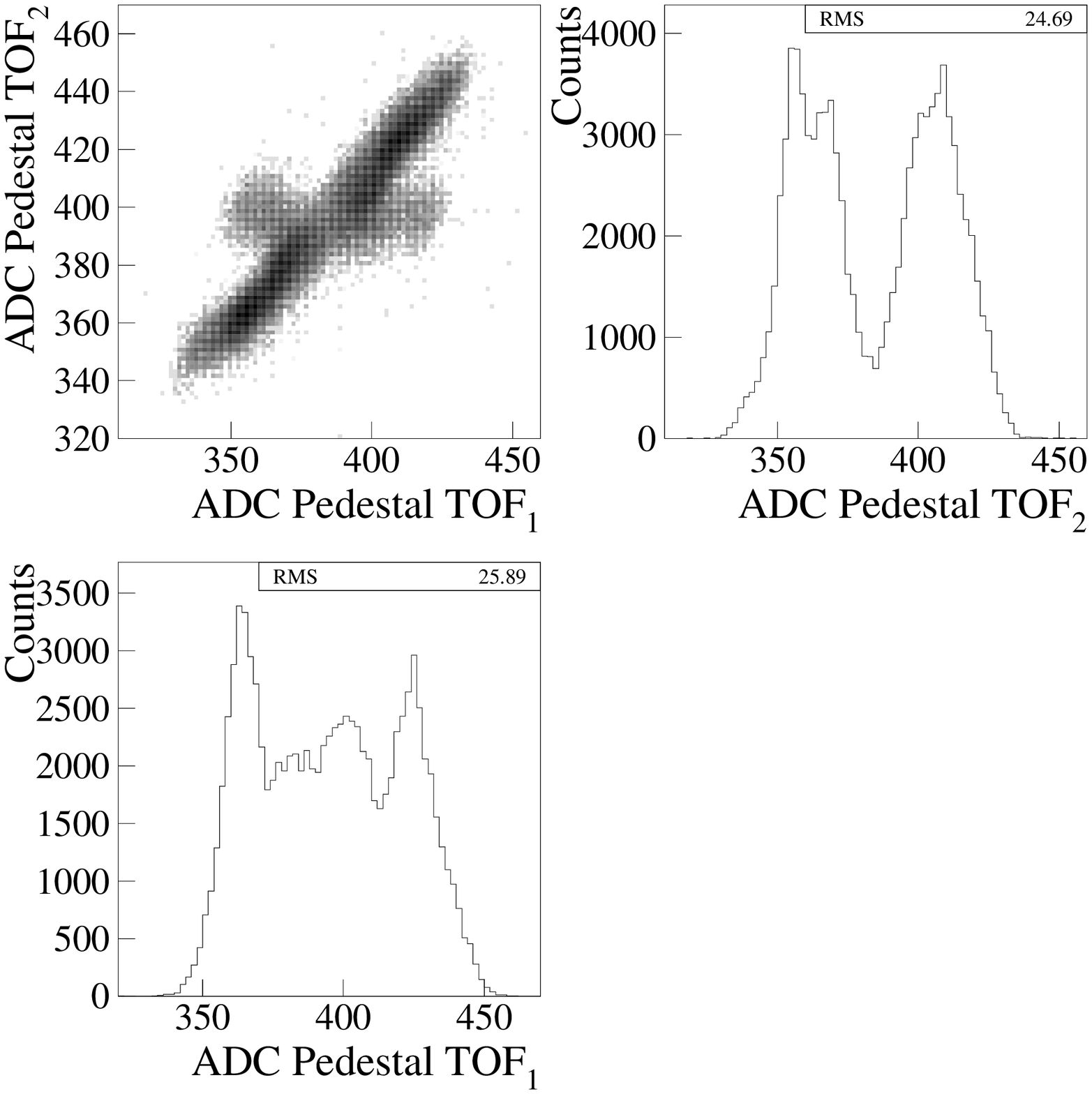 hscale=0.4 vscale=0.4} 
\vskip -0.2cm 
\caption{Two typical ADC pedestal distributions for two channels in
the E896-TOF System during the Au98 run (upper right and lower left),
and their correlation in two dimensions (upper left). The axes out of
the page are logarithmic.\label{fig:peds}}
\end{figure}

All of the $\sim$600 ADC pedestals in five different detector systems
in E896 are, in some way, highly correlated during full running
conditions. An example is shown in Fig. \ref{fig:peds} for two PMTs
attached to two different E896 TOF slats. The upper right and lower
left frames in Fig. \ref{fig:peds} indicate that the ADC pedestals for
these two channels have variances on the order of 40 channels, which
is an order of magnitude larger than would be expected in the absence of
correlated noise. 

To recover the needed ADC resolution, one must measure the
contribution to the ADC values arising solely from the correlated noise
component in each experimental event. One method is to add one or more
``blackened PMTs" (bPMTs) to the detector, and digitizing these in
ADCs exactly as if these bPMTs were attached to active elements.
Such blackened PMTs are optically isolated by definition, but as
they are mechanically and electrically part of the detector, the
ADC values measured are, up to (time-independent) offsets,
exactly the correlated noise contribution event by event.

\begin{figure}[htb]
\vspace{4.0cm} 
\hspace{2.5cm}
\special{epsf=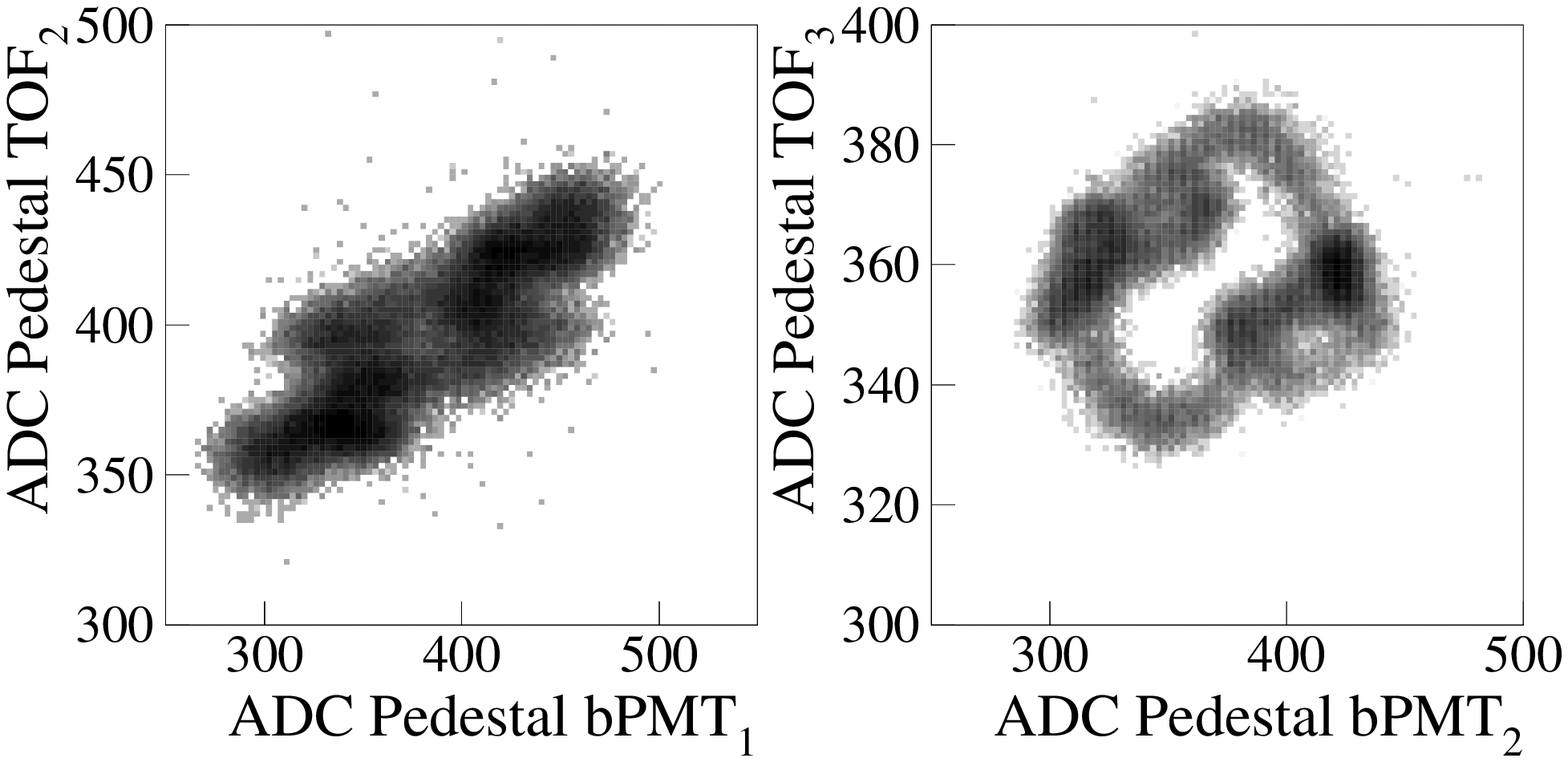 hscale=0.4 vscale=0.4} 
\vskip -0.2cm 
\caption{Correlations of the ADC pedestals from two active channels
(each ordinate) and two bPMT channels (each abscissa) in the TOF
system in pedestal-only events. The axes out of the page are
logarithmic.\label{fig:bpmt}}
\end{figure}

If the correlation between the ADC pedestal from an active channel and
that from a bPMT channel is sufficiently strong and single valued
({\it i.e.} something like the left frame in Figure \ref{fig:bpmt},
the ADC values in active channels can be corrected with some
efficiency for the correlated noise contribution using the value of
the bPMT in each experiment event. One makes two passes through the
experimental data. First one records the correlation between the ADC
values for the active channel versus the ADC values from a bPMT, in
only those events when the active channel is known not to be struck by
a particle. Once this dependence versus one (or more) bPMTs is known
for each active channel, in the second pass through the data the bPMT
correction can be applied to the active channels in all events,
independent of whether the active channel was struck by a particle or
not.

However, there is a crucial shortcoming of such bPMT-based approaches.
A complete offline correction is possible only if this
correlation is indeed strong and single-valued. That is, the blackened
PMT must sit on the same local ground as the PMT to be
corrected. In electrically simple experiments, {\it i.e.} if there is
only one or two variants of the (dirty) electrical ground seen by
different detector elements, a handful of blackened PMTs could provide
all of the information needed for a sufficient correction for
correlated noise. In more complicated experiments, {\it i.e.} anything
on a heavy-ion beam-line, it is possible that
innumerable different local grounds exist with different shapes,
amplitudes, and phases. An example of a detector channel that cannot
efficiently be corrected using a particular bPMT is seen in the
right frame of Figure \ref{fig:bpmt}.

\begin{figure}[htb]
\vspace{9cm} 
\hspace{1cm}
\special{epsf=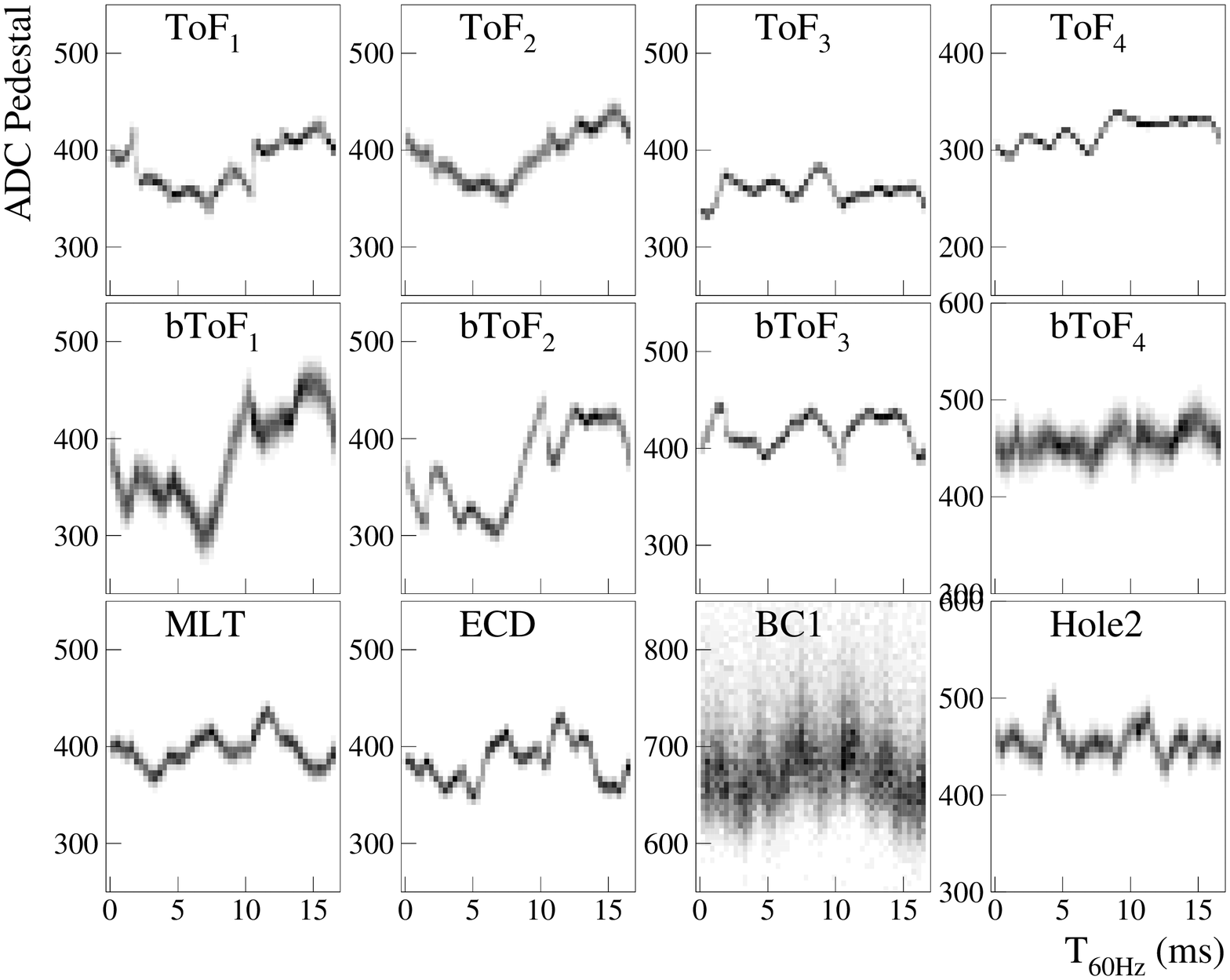 hscale=0.45 vscale=0.45} 
\vskip -0.2cm 
\caption{The ADC pedestals from a sampling of PMT-equipped detectors
in E896, as labelled in each frame, versus a particular event time
described in the text. The axes out of the page are logarithmic.\label{fig:profiles2}}
\end{figure}

Shown in Figure \ref{fig:profiles2} is the correlation of ADC
pedestals from a number of detectors in E896 versus a particular time
recorded for each experimental event and inserted into the main data
stream. This time is obtained from a clock, described below, that
resets at 60 Hz in active synchronization with the experimental AC
power. These plots can be considered oscilloscope traces for ADC
pedestals where the horizontal axis is \sixteenms in total, and the
``trigger" is by definition always at exactly the same phase relative
to the experimental AC power line. Clear signatures of correlated
noise are in fact seen in {\it all} PMT-equipped channels in this
experiment during full beam-on running conditions. It is also clear
that there are many variants of the (dirty) ground local to the
different subdetectors in E896, and indeed differing local grounds
also exist between different channels in a given subdetector.

If one wanted to combat the noise problem in E896 with blackened PMTs,
one would need {\it many} of them, and the time and tenacity to find {\it all}
of the different local grounds at work in the experiment. A more
universal and technically simpler correction is possible given the
availability of a single number for each experimental event. This
number is the time the event occurred as measured within \sixteenms
intervals that are actively synchronized with the experimental AC
power, {\it i.e.} the time defining the horizontal axis of Fig.
\ref{fig:profiles2}. The custom electronics that we developed that
allows this number to be trivially inserted into any experimental data
stream for each experimental event is described in the next Section.

\section{The Ramp\label{sec:blackbox}}

Given the time that each event occurs, measured relative
to \sixteenms intervals that are actively synchronized with the
experimental AC power, one can correct completely for correlated noise
offline. In experiments in the U.S., the line frequency is 60 Hz.
One thus can use an AC line-synchronized 60 Hz pulser and a
latching scaler to provide such clock information to a data
acquisition system.\cite{ref:helio} We describe here another approach.
The present approach was developed primarily because it is more
suitable for integration into the E896 data stream than the
pulser+latching scaler approach, although the circuit we built is also
smaller, considerably cheaper, and more generally portable. The
present device can be made part of any experiment that has one spare
AC power socket and one spare ADC channel.

The circuit generates a precision negative-voltage sawtooth waveform
which resets from V$_{max}$ to V$_{min}$ in active synchronization
with the experimental AC power line. The output is sent to a
spare channel of the ADCs used to digitize the detector signals. When
an event occurs, the ADC channel connected to the ramp is
thus gated at the same absolute time as the detector channel ADCs are
gated. Experimental gates are typically 100-200 ns wide,
which is a factor of 10$^5$ shorter than the \sixteenms period of the AC line.
This implies the gating of the ADC channel connected to the ramp 
digitizes a very thin vertical slice, so the pulse area
measured by the ramp ADC is thus effectively just a measurement of
the instantaneous value of the ramp voltage at the time the experimental event
occurs. 

By construction, the ramp voltage so measured is related
linearly to the time within actively AC line-synchronized \sixteenms
intervals. The depiction of an ADC pedestal, from detectors anywhere
in the experiment, versus the (single) ramp ADC value in the same
event thus measures exactly, event by event, the time dependence of
the dirty experimental ground local to each detector channel. This
allows straightforward and complete offline corrections (examples
described below), no matter how many different local grounds exist in
the experiment. As there is by definition a ramp value for every
experimental event, the correction (described in the next section) is
$\sim$100\% efficient in practice.

\begin{figure}[htb]
\vspace{10.1cm} 
\hspace{0cm}
\special{epsf=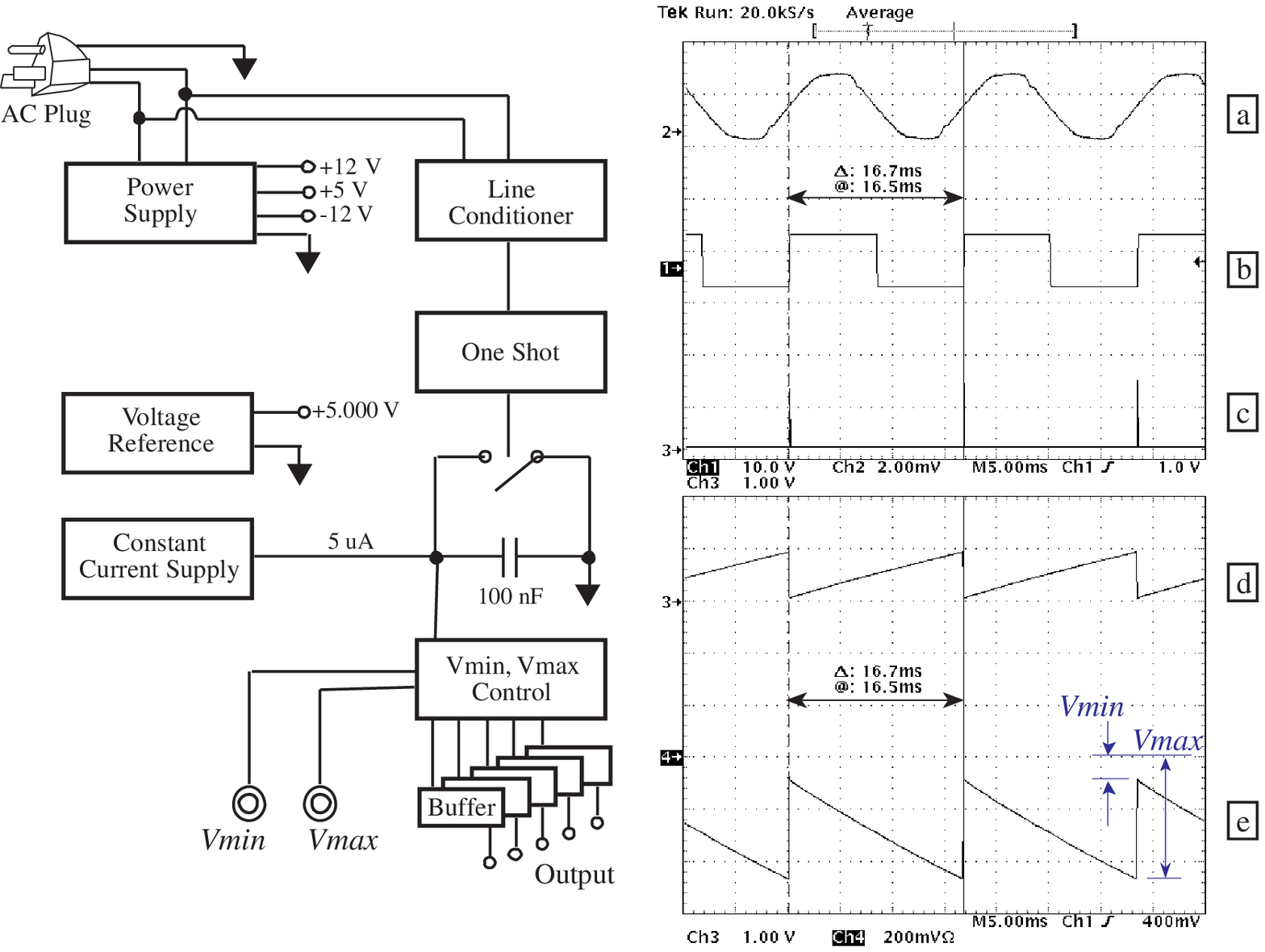 hscale=0.78 vscale=0.78} 
\vskip -0.2cm 
\caption{On the left side, a schematic depiction of the
present synchronized clock circuit, and on the right are
oscilloscope traces for the signals at various stages
of the signal processing.\label{fig:rampdiagram}}
\end{figure}

The circuit that was built in early 1998 and used in the E896
1998 Au run is shown schematically in Figure \ref{fig:rampdiagram}.
The circuit is connected to the experimental AC power using a
US-standard three prong plug. The AC voltage is shown, attenuated
by 100 dB, as trace (a) in this figure. The neutral and phase
lines from the AC plug are sent to a power supply, which provides
the low voltages used by the circuit, and to a line conditioner.
The line conditioner searches for the time at which the
experimental AC power crosses zero Volts from below, and
generates a bipolar square wave in phase with this time, and
hence with the AC power cycle, as seen in trace (b). The bipolar
square wave is sent to a ``one-shot," which controls a switch.
The leading edges of the one-shot output, shown in trace (c), open
the switch and the falling edges close the switch.

A voltage reference and a constant current supply continuously
add charge to a 100 nF capacitor (trace d), and by a related
amount to the rest of the circuit. This produces the precision
voltage ramp, shown in trace (e). When the next 60 Hz cycle
begins, {\it i.e.} on a leading edge of trace (b), the one shot
closes the switch and discharges the capacitor. When the circuit
senses the capacitor is fully discharged, which is approximately
3 $\mu$s later, the one-shot reopens the switch and the output
(trace e) returns to V$_{\rm min}$. Current then begins flowing into the
capacitor again, charging it and producing an output voltage that
again becomes more negative in precise linearity with the time
within the present AC cycle. 

The synchronization is active, occurring independently for each cycle
of the AC power. Thus, in principle there is nothing in the present
circuit that makes it specific to 60Hz line frequencies. The device
should work without modification for others, such as are used in
experiments abroad.

The discharging of the capacitor implies a ``dead time" of the circuit
of $\sim$3 $\mu$s/\sixteenms, or about 0.02\%. The current out of the
constant current supply is generated against a 5.000V voltage
reference, which holds the voltage on the capacitor precise to one
part in a few thousand. This implies that the circuit provides the
time within AC line-synchronized \sixteenms periods with a resolution
of a few microseconds. Such a time resolution is well better than
necessary to correct for typical correlated noise levels - typically
one needs sufficient statistical certainty in only 50-100 bins of the
60Hz time in order to perform a complete correction (as shown below).

The values of V$_{\rm min}$ and V$_{\rm max}$ are adjustable via
externally mounted potentiometers. This allows the device to be
trivially implemented in other experiments, which may use ADC
gates of different widths or ADC modules with different
full-scale ranges or charge conversion factors. The present
circuit uses buffers to produce five copies of the ramp signal,
which were each sent to a spare channel in five different TOF ADC
modules in the E896 TOF system. In the end only one output
channel was actually needed, as the circuit proved to be very
stable and its data easy to understand.

\begin{figure}[htb]
\vspace{7.5cm} 
\hspace{1.cm}
\special{epsf=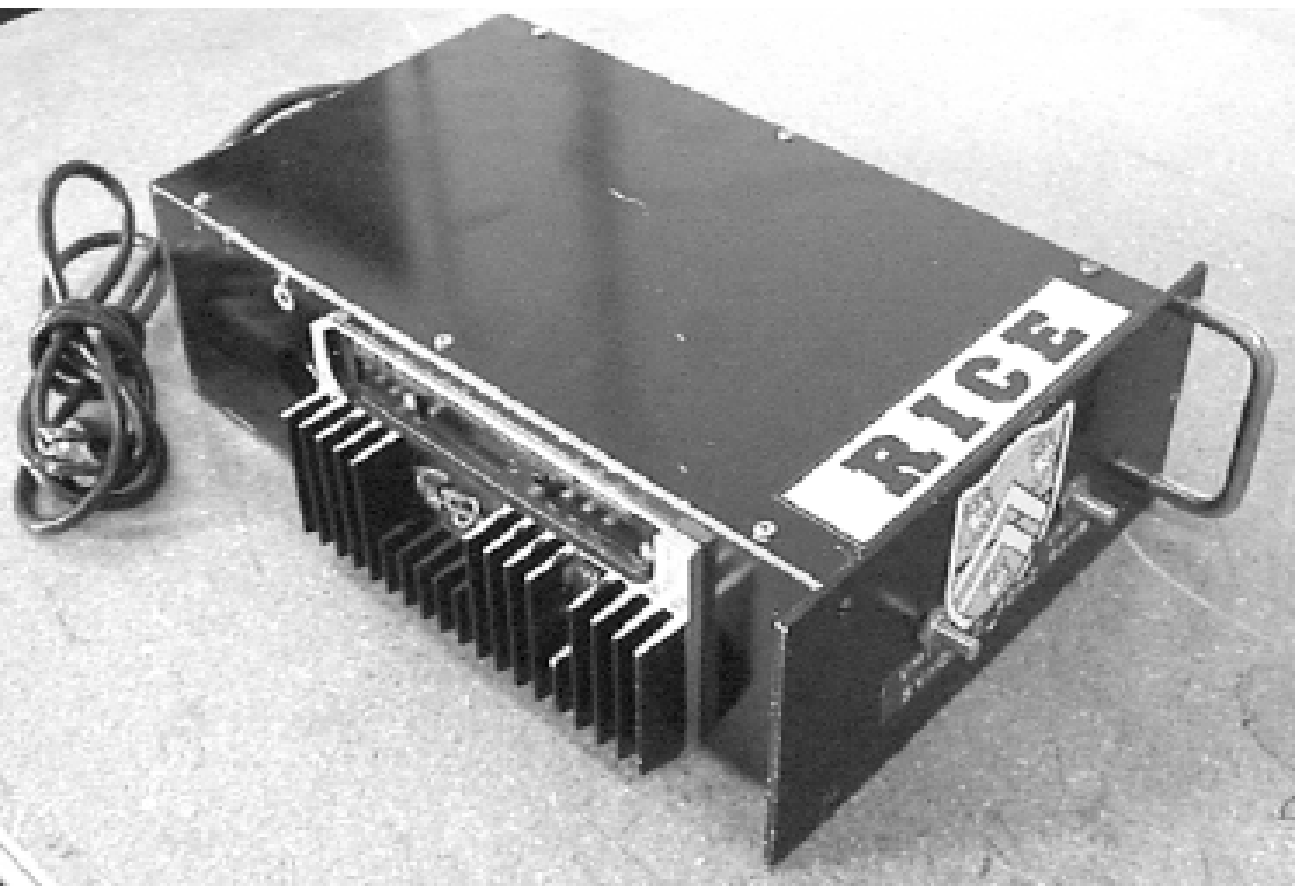 hscale=0.87 vscale=0.87} 
\vskip -0.2cm 
\caption{A side and front view of ramp generator box. Visible
on the front panel are the two dials for the V$_{\rm min}$
and V$_{\rm max}$ control, and on the side are cooling vanes.
Not visible on the back of the box are the five BNC connectors 
for the output of the ramp signals. \label{fig:ramp}}
\end{figure}

The circuit is packaged in a small metal box with external signal
connectors and the pots for the V$_{min}$ and V$_{max}$ control,
which is shown in Fig. \ref{fig:ramp}. The cost was about \$100 
in total.

\section{The Performance\label{sec:results}}

\begin{figure}[htb]
\vspace{6.9cm} 
\hspace{3cm}
\special{epsf=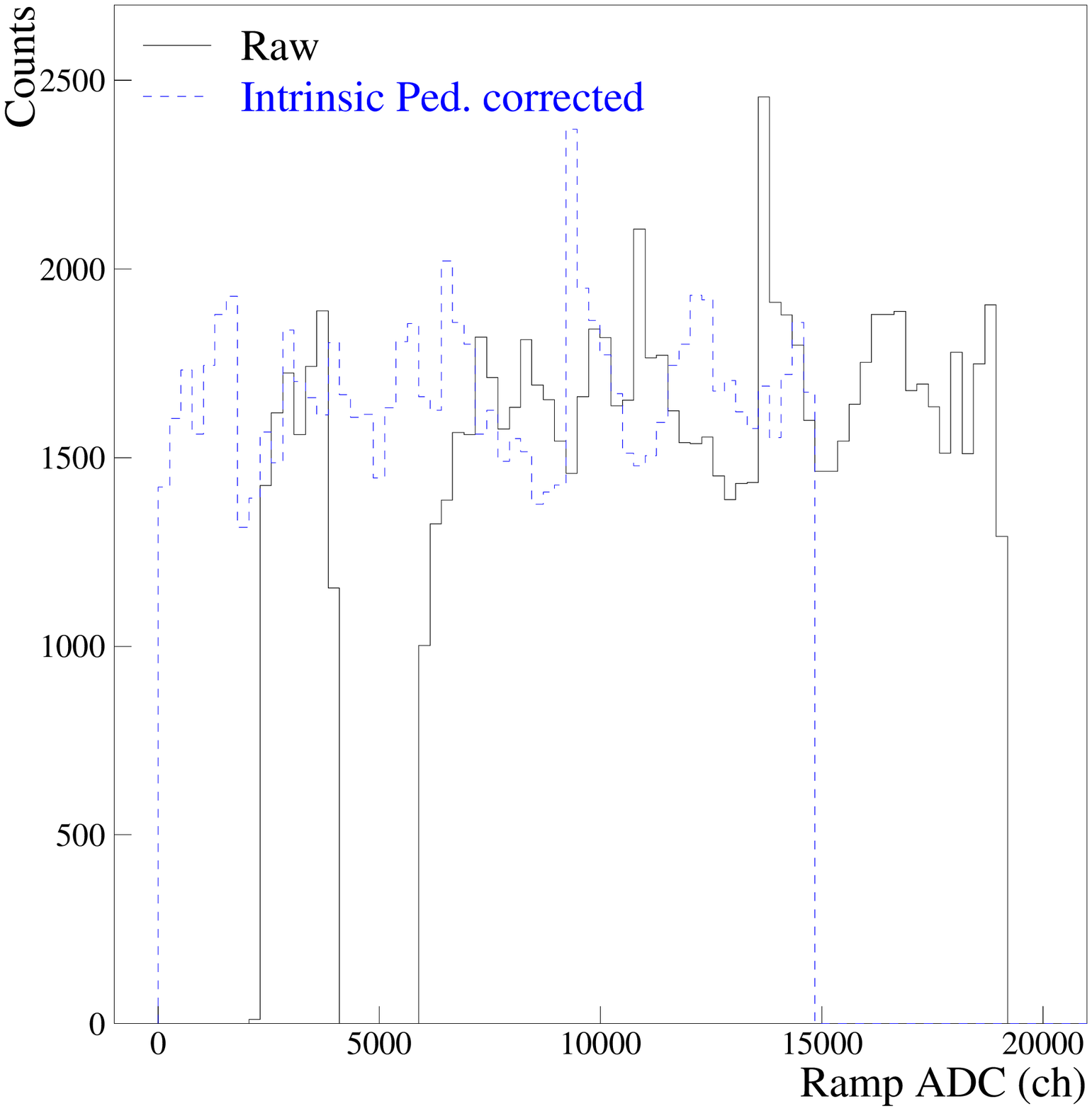 hscale=0.35 vscale=0.35} 
\vskip -0.2cm 
\caption{An example from one run of the event-inclusive distribution
of ADC values for the channel connected to the present ramp
circuit. The raw ramp ADC distribution is shown in the solid
line, while the dotted line shows the ramp ADC distribution following the 
subtraction of the two intrinsic pedestals per channel in the
(dual-range) LRS 1885F ADCs.\label{fig:ramp_hist}}
\end{figure}

The event-inclusive distribution of ADC values for the ADC channel
connected to the present ramp circuit should, except for
spill-structure or other known event trigger time correlations, be
essentially flat. Such a distribution for one data-taking run from the
E896 Au98 run is shown in Figure \ref{fig:ramp_hist}. The raw ramp ADC
distribution is shown in the solid histogram. The dashed histogram
depicts the same ramp ADC distribution following the subtraction of
the two intrinsic pedestals in the (dual range) LRS 1885F ADCs. This
subtraction was done following a measurement of these two intrinsic
pedestals for each ADC channel using calibration logic internal to the
ADC and a Tcl/Tk script.\cite{ref:salvo}

\begin{figure}[htb]
\vspace{9cm} 
\hspace{1cm}
\special{epsf=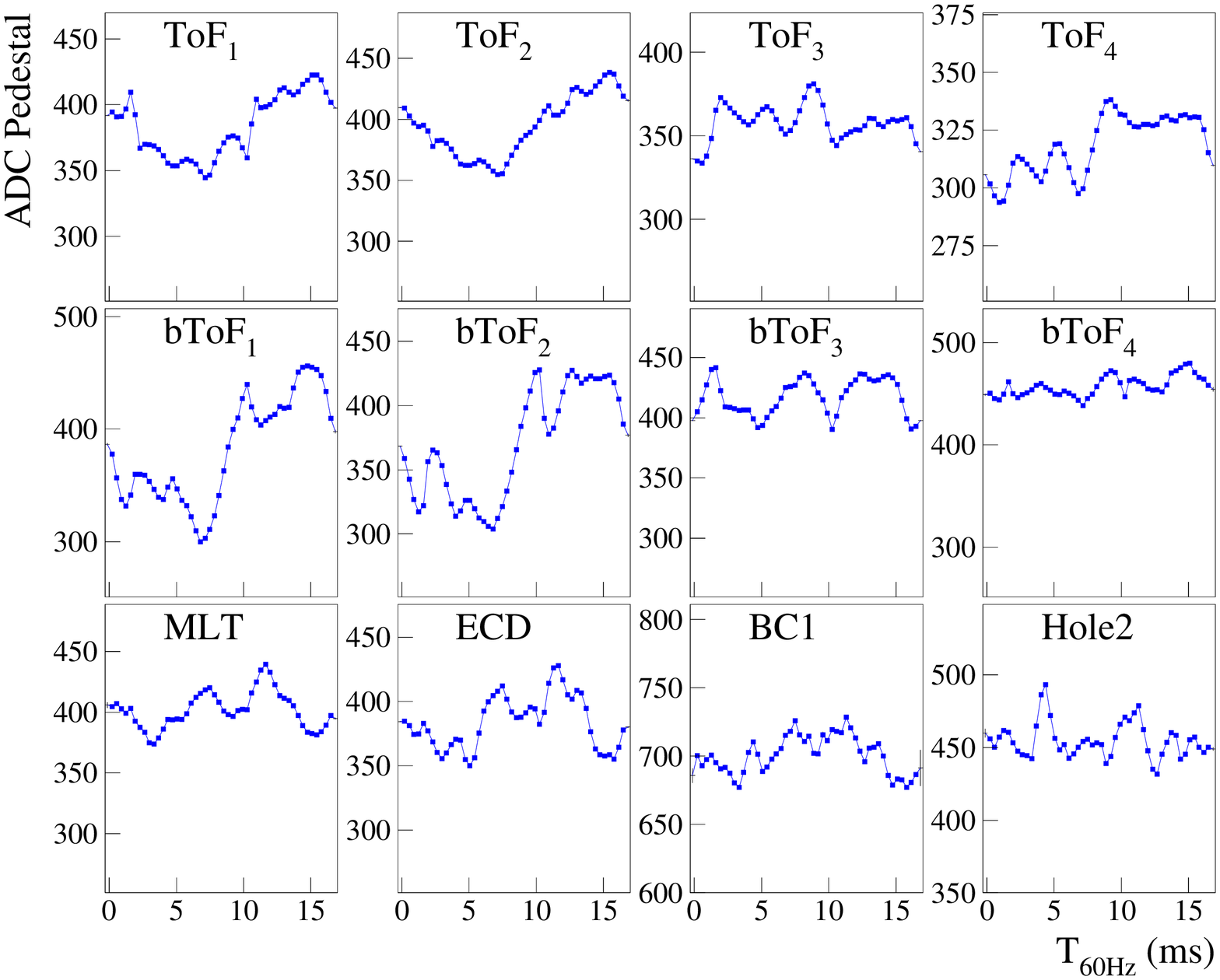 hscale=0.45 vscale=0.45} 
\vskip -0.2cm 
\caption{Profile histograms of the correlation of the average ADC
pedestals in a single experimental run versus the time line
line-synchronized \sixteenms intervals for a sampling of PMT-equipped
detector channels in E896.
\label{fig:profiles}}
\end{figure}

Based on the dashed histogram in Figure \ref{fig:ramp_hist}, the times
$t_0$ at which new 60 Hz cycles begin (and the ramp resets to the
voltage $V_{min}$) are identified as those events with a
pedestal-subtracted ramp ADC value of 0. The time \sixteenms later is
identified as an pedestal-subtracted ramp ADC value of $\sim$15,000.
Thus the event time, $T_{60Hz}$, in milliseconds relative to the AC
cycle is then computed from the intrinsic pedestal-subtracted ramp ADC
values in each event, $ADC_{ramp}$, simply from $T_{60Hz}$ $=$ (16.67
ms)*$ADC_{ramp}$/$15000$. This is how the horizontal axis of Figures
\ref{fig:profiles2} and \ref{fig:profiles} was obtained. It should be
noted that, as the digitization of the ramp voltage is performed
entirely in the counting house, and no PMTs are involved, there is no
correlated noise contribution in the ramp ADC values themselves. This
allows the ramp ADC values to be applied directly for the noise
correction of all of the PMT-equipped detector channels in the
experiment.

During the 1998 Au run, hundreds of data runs were recorded to tape in
E896, and each was approximately 20 minutes long ($\sim$200k
events/data run). This is well matched to the time scale of roughly
hours over which the correlated noise line shapes
for a given detector channel slowly drift. Thus, the
recording the pedestal profiles for each of the $\sim$600 ADC
pedestals in E896 once per data run allows sufficiently accurate
corrections. 

A correction based on the present device involves two passes through the
experimental data. In the first, the correlated noise ``profiles" 
versus the ramp ADC are collected and saved for each PMT in the experiment.
Shown in Figure \ref{fig:profiles} are such profiles, {\it
i.e.} the ordinates are average values of an ADC pedestal for all
events in a particular 60Hz time bin (abscissa), for the same sampling of
E896 PMTs across different detector systems that were shown in Fig.
\ref{fig:profiles2}. 

In any one of these frames, the average ADC
values are the same at the extreme values, indicating continuity and
the appropriateness of the present 60 Hz clock-based corrections. It is also
interesting to notice in Figure \ref{fig:profiles} (or Figure
\ref{fig:profiles2}) that frequency modes that are some multiple of 60
Hz are clearly also present. An analysis of the frequency spectra
obtained from Fourier transforms of these line-shapes for each of the
$\sim$380 PMTs in the E896 TOF system will be presented in the next
section.

In all subsequent passes through the experimental data, the value of the
ramp ADC in each experimental event is then used to look up $\sim$600
values from the $\sim$600 profile histograms recorded in the previous
pass. For each detector channel in this event the appropriate
60Hz-time-dependent pedestal is subtracted from the detector channel's
ADC value, whether or not there is a ``hit" in this particular
detector channel in this event. The ADC pedestal is then, in software,
completely restored to the much-tighter distribution expected in the
absence of the correlated noise. The resolution for ``hits" may also
be improved.

\begin{figure}[htb]
\vspace{8cm} 
\hspace{3cm}
\special{epsf=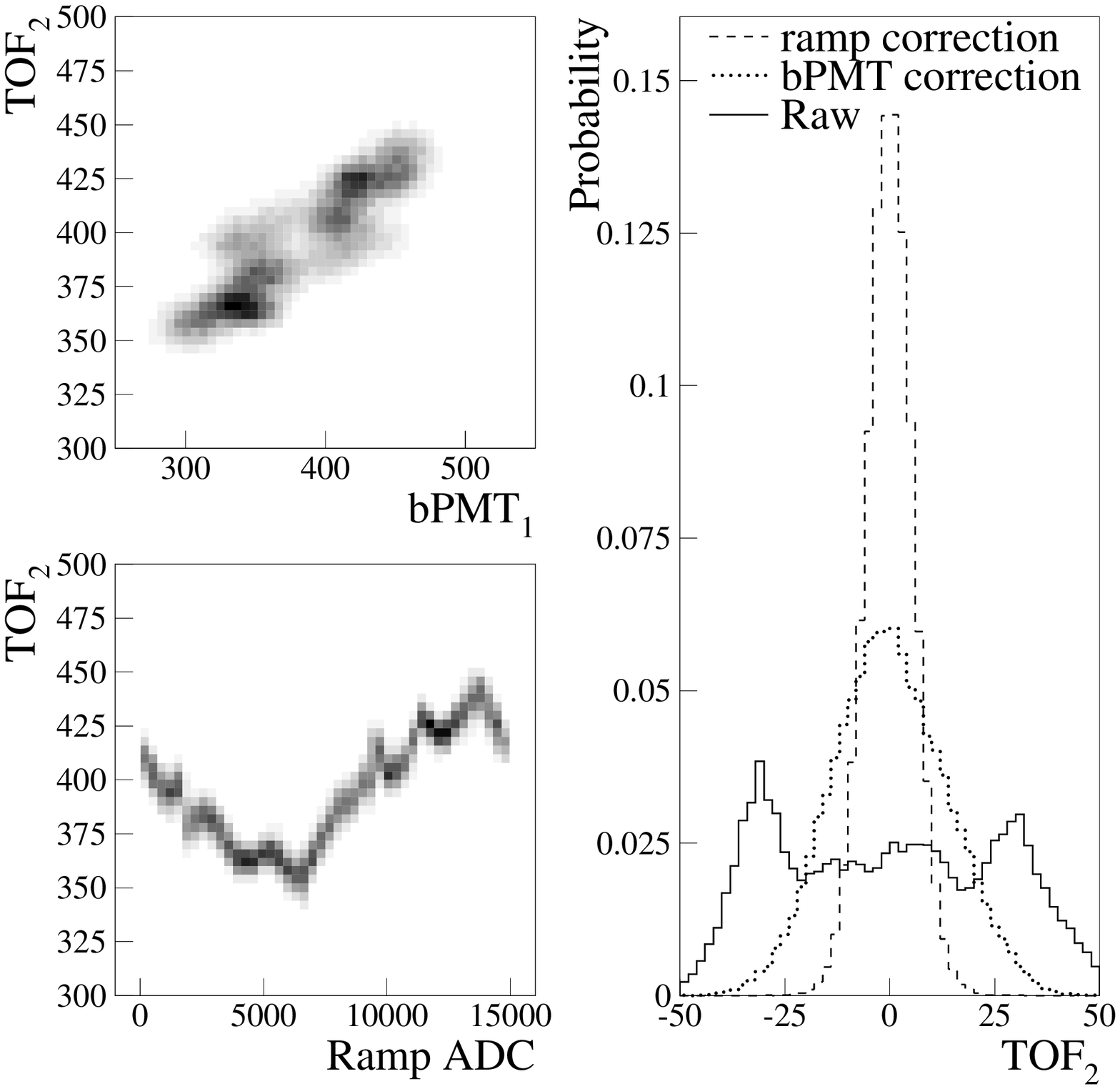 hscale=0.4 vscale=0.4} 
\vskip -0.2cm 
\caption{The comparison of correlated noise corrections based on bPMTs
and the present clock circuit for the same two channels (one from the
TOF and one a TOF bPMT) shown in the left frame of Figure
\ref{fig:bpmt}. The axis out of the page in the left frames
is linear.
\label{fig:correct1}}
\end{figure}
\begin{figure}[htb]
\vspace{8cm} 
\hspace{3cm}
\special{epsf=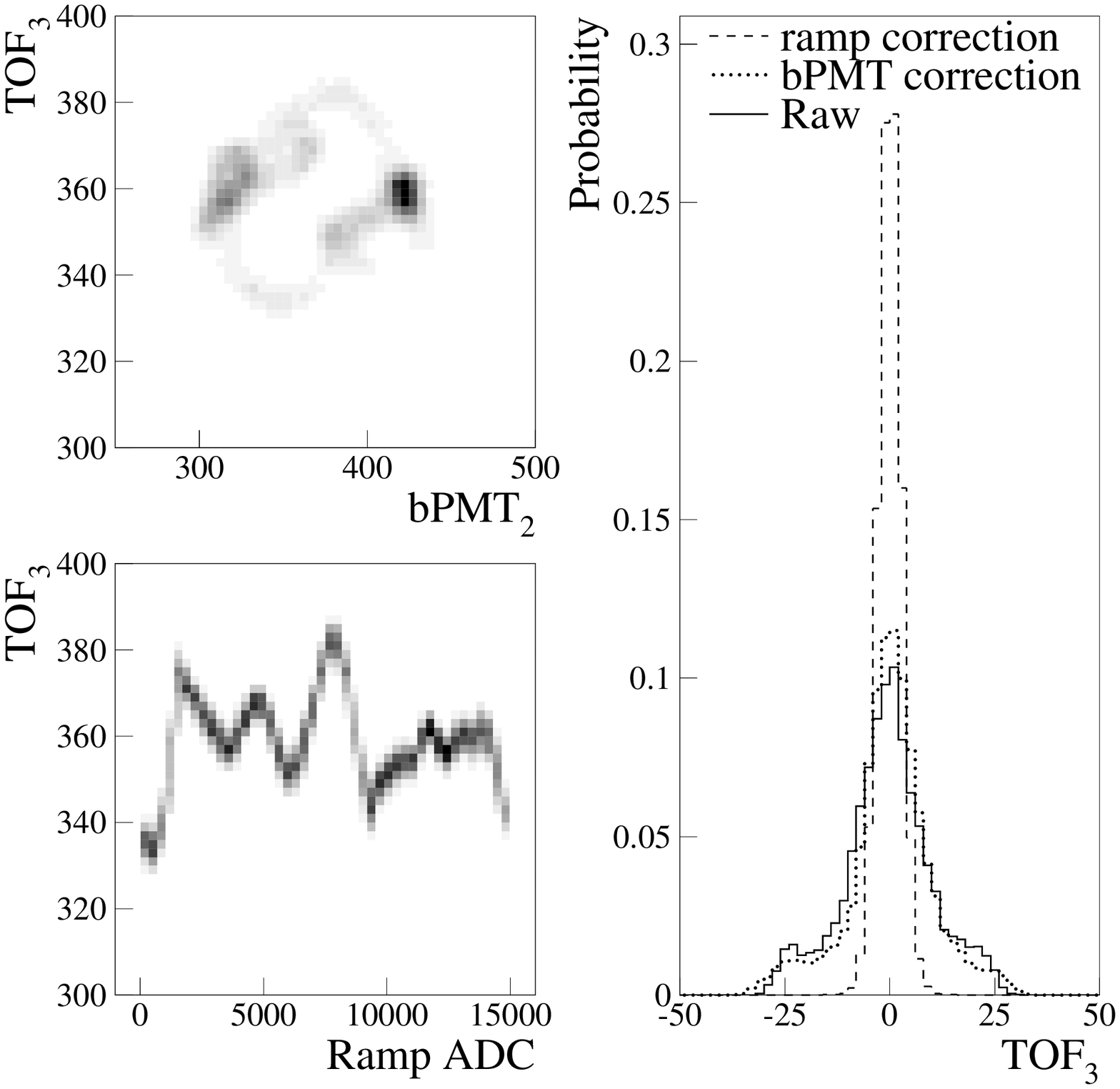 hscale=0.4 vscale=0.4} 
\vskip -0.2cm 
\caption{The same as Figure \ref{fig:correct1}, but for the same
channels shown in the right frame of Figure \ref{fig:bpmt}.\label{fig:correct2}}
\end{figure}

Examples of the dramatic improvement to correlated noise problems in
these data is shown in Figures \ref{fig:correct1} and
\ref{fig:correct2}. In either figure, the upper left frame is, for
reference, a correlation between a TOF ADC pedestal and a TOF bPMT
pedestal. This TOF pedestal is plotted versus the (dual pedestal
subtracted) ramp ADC value, {\it i.e.} the time in AC line synchronized
\sixteenms intervals, in the lower left frame. The comparison of the
two corrections, one based on this bPMT ADC and the other based on the
ramp ADC, is shown in the right frame. The raw pedestal is shown as
the solid histogram, the pedestal following the bPMT correction is the
dotted histogram, while the pedestal following the ramp ADC correction
is the dashed histogram.

One observes in the right frame of Figure \ref{fig:correct1} that the
bPMT correction leads to only a partial correction, while in the right
frame of Figure \ref{fig:correct2}, the bPMT provides no significant
correction at all. This is due to the poor correlations seen in the
upper left frames. However, as seen in both of the lower left frames,
a far stronger correlation exists between the TOF pedestals and the
ramp ADC values, allowing an efficient and complete correction.
According to the two right frames of these Figures, the ramp
correction restores the ADC pedestals to perfect Gaussian's with a
variance of 3-5 channels - as would be expected in the absence of correlated
noise.

\section{Spectral Analysis\label{sec:fourier}}

That the time dependence of the value of any E896 ADC pedestal is
manifestly periodic at 60 Hz is apparent from the tight correlation,
and the continuity at the endpoints, of the profiles shown in Figures
\ref{fig:profiles2} and \ref{fig:profiles}. This makes no statement on
higher frequency modes though. As the correlated noise is absent when
only the experiment itself is fully powered, but is present during
full beam-on running, one suspects a contributor to correlated noise
at the AGS is the massive power supplies connected to the numerous
beam-line magnets. These regulate at 720 Hz in multiples of 60, so one
might expect that frequency modes that are multiples of 60 Hz exist in
the line-synchronized time dependence of the ADC pedestals. To
investigate this possibility, the measured values of the ADC pedestals
versus the time from our 60 Hz clock are Fourier analyzed to extract
frequency spectra.

For each of the $\sim$384 channels of the TOF system, the ADC pedestal
versus ramp ADC profile, {\it e.g.} Figure \ref{fig:profiles}, was
measured using the data from a single data run. For each profile, an
array of dimension 2$^{11}$ was filled with 60 copies of this profile
placed end to end in this array. This array thus contains an average
PMT pedestal as a function of the time in 2$^{11}$ bins that spans a
total period of 1 second. The Fourier transform of this array then
contains the weights for frequencies measured in Hertz.

\begin{figure}[htb]
\vspace{5.8cm} 
\hspace{1cm}
\special{epsf=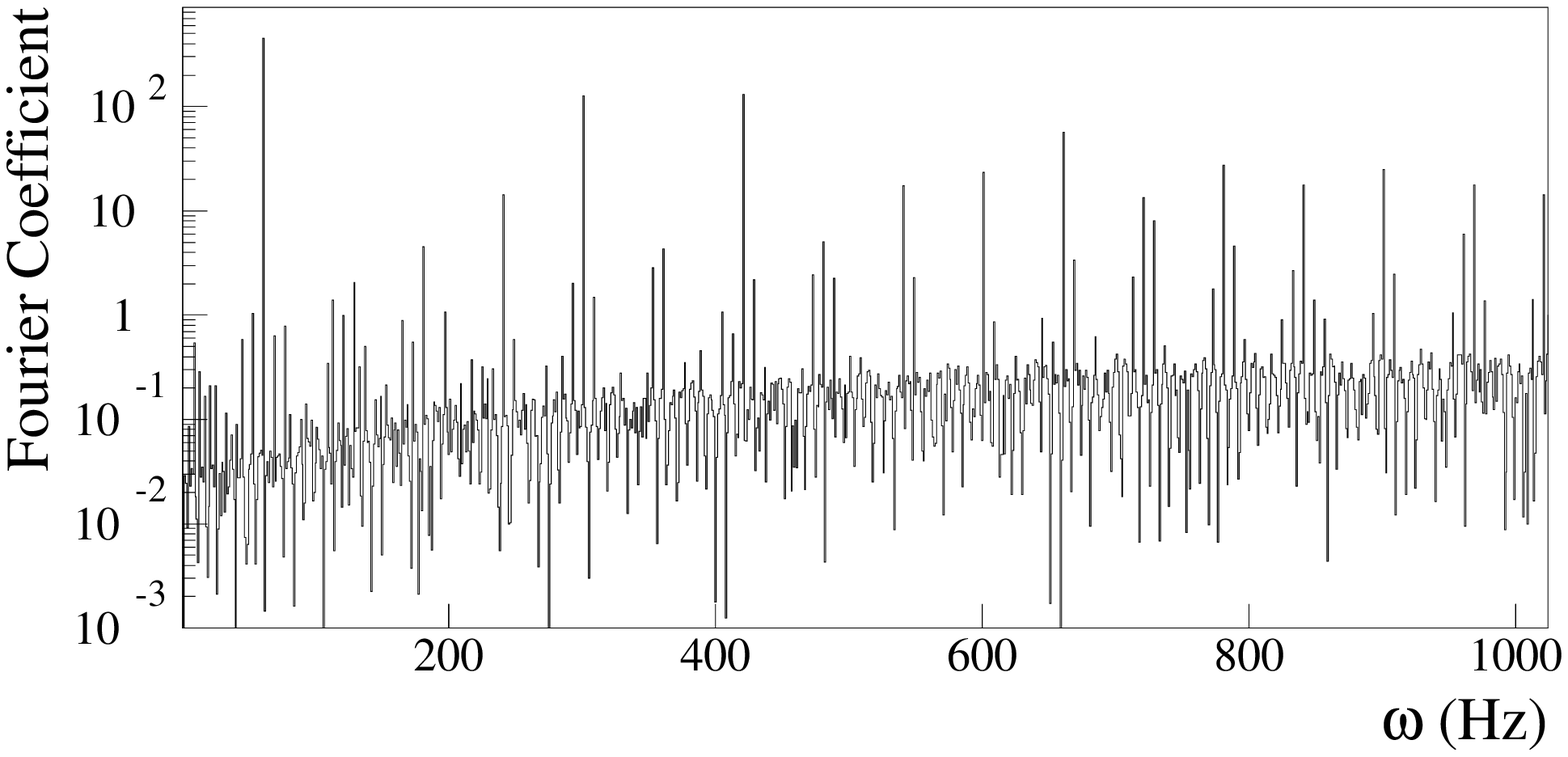 hscale=0.6 vscale=0.6} 
\vskip -0.2cm 
\caption{The Fourier coefficients versus frequency, $\omega$, in Hertz
for the correlated noise profile for a particular TOF detector
channel.\label{fig:freq}}
\end{figure}

A typical frequency spectrum is shown in Figure \ref{fig:freq}. The
weight for the zero-frequency mode, which is related to the (time
independent) mean value of the pedestal, is suppressed in this
histogram. The dominant frequency modes are identified as the bins in
this histogram with the largest absolute values of the Fourier
coefficient in this bin. The statistical nature of the mean ADC values
versus the 60Hz time used as input to the Fourier transforms results
in a finite weight for all possible frequency modes. For the TOF
channel shown in this Figure, the dominant mode is 60 Hz, although
there is significant weight in modes at 300 Hz and 420 Hz.

\begin{figure}[htb]
\vspace{6cm} 
\hspace{1cm}
\special{epsf=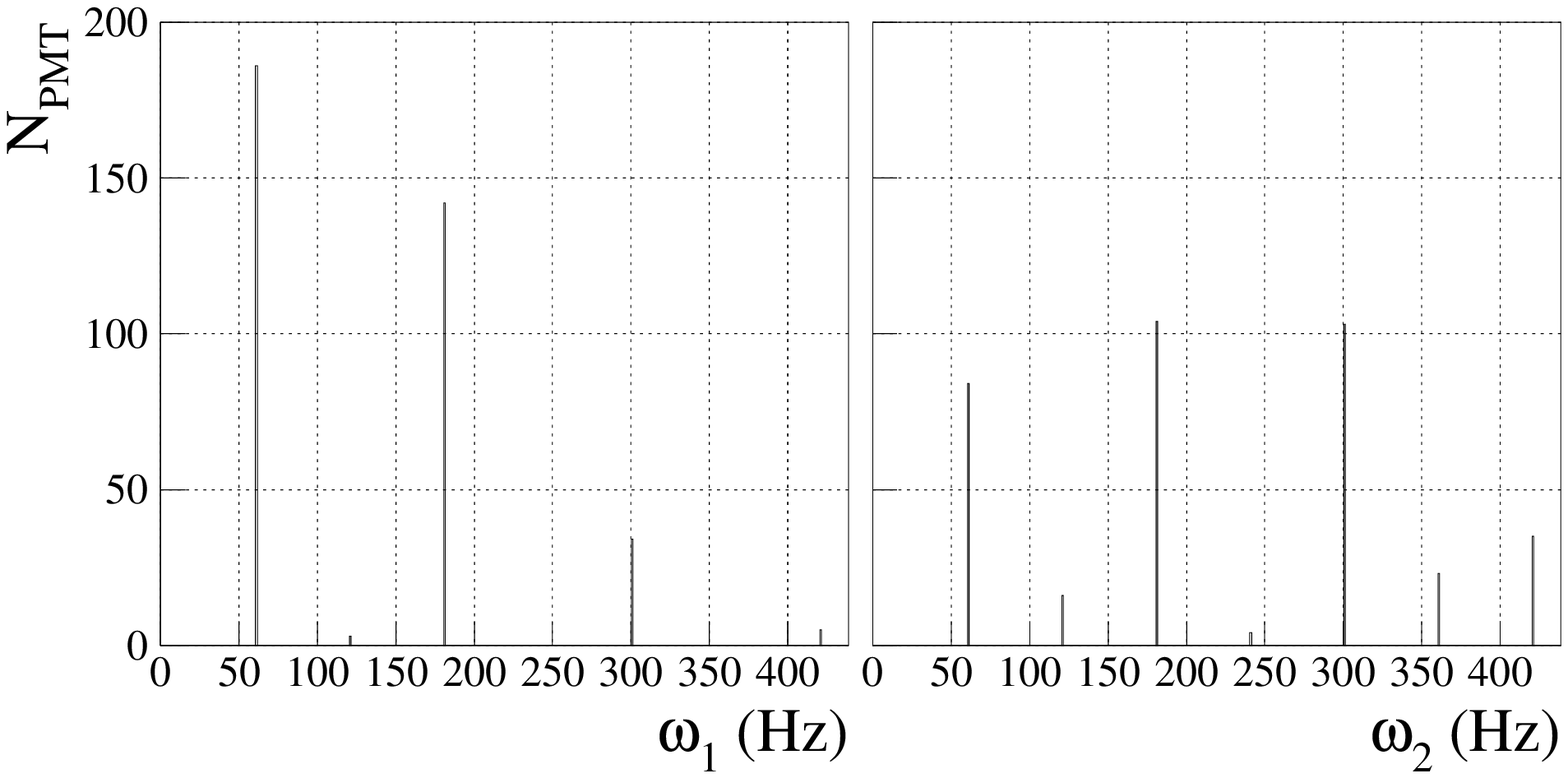 hscale=0.6 vscale=0.6} 
\vskip -0.2cm 
\caption{On the ordinates are histograms of the number of E896 TOF
channels that have particular values of the dominant frequency
(abscissa, left frame) and sub-dominant frequency (abscissa, right
frame).\label{fig:modes}}
\end{figure}

For each of the 384 channels of the E896 TOF system and in one run,
the first and second largest Fourier coefficients were located for
each channel using this procedure. The histogram of the number of TOF
detector channels having a particular value of the frequency with the
largest(second largest) Fourier coefficient is shown in the
left(right) frame of Figure \ref{fig:modes}. Roughly half of the
channels of the E896 TOF system have pedestals that depend on time
with the dominant frequency mode being exactly 60 Hz. However, the
other half of the TOF channels have ADC pedestals that depend on time
dominantly at frequencies that are 3, 5, or 7 times 60 Hz. Even
multiples of any significance were not seen in the dominant frequency
mode (left frame), although there are channels for which even
multiples contribute to the sub-dominant mode (right frame).

It is interesting to note that these is a certain amount of
correlation between the dominant frequency mode of the time dependence
of the ADC pedestal for a given channel and the location of this
channel in the apparatus. In principle this information could be used
to provide more clues as to the exact sources of the correlated noise.
In practice, however, this is not necessary, as the offline 60 Hz
clock-based corrections completely solve the problem anyway.

\section{Summary\label{sec:summary}}

A small, highly portable, and inexpensive custom circuit was developed
to provide the information needed for an efficient and complete
offline correction for correlated noise in all of the $\sim$600 PMT-equipped
detector channels in BNL-AGS Experiment 896. The circuit generates a
precision voltage ramp that resets in active synchronization with the
AC power, and this ramp voltage is simply digitized in a spare
ADC channel to provide line synchronized clock information for
every experimental event. The variances of the experimental ADC
pedestals following the ``ramp ADC" correction were reduced from, in
some cases, tens of channels to the values of 3-5 channels expected in
the absence of the correlated noise. Experimentally, the correlated noise 
profiles for each and every PMT was fundamentally periodic and repeating
at 60Hz, yet a considerable number of channels indicate predominant
frequencies of odd multiples of 60 Hz, primarily 180 Hz and 300 Hz.

\ack
We thank the BNL-AGS E896 Collaboration for the use of preliminary ADC
pedestal data from the Au98 run that was discussed above. We gratefully
acknowledge funding from the US Department of Energy under Grant No.
DE-FG03-\-96ER40772, as well as helpful comments from H. Takai, S.
Costa, C. Nociforo, G.S. Mutchler and E.D. Platner.


\end{document}